# Metalloboranes from first-principles calculations: A candidate for high-density hydrogen storage

A. R. Akbarzadeh, D. Vrinceanu, and C.J. Tymczak

*Department of Physics, Texas Southern University,
Houston, Texas 77004, USA*

**Abstract**

Using first principles calculations, we show the high hydrogen storage capacity of a new class of compounds, metalloboranes. Metalloboranes are transition metal (TM) and borane compounds that obey a novel-bonding scheme. We have found that the transition metal atoms can bind up to 10 $H_2$-molecules with an average binding energy of 30 kJ/mole of $H_2$, which lies favorably within the reversible adsorption range. Among the first row TM atoms, Sc and Ti are found to be the optimum in maximizing the $H_2$ storage on the metalloborane cluster. Additionally, being ionically bonded to the borane molecule, the TMs do not suffer from the aggregation problem, which has been the biggest hurdle for the success of TM-decorated graphitic materials for hydrogen storage. Furthermore, since the borane 6-atom ring has identical bonding properties as carbon rings it is possible to link the metalloboranes into metal organic frameworks (MOF's), which are thus able to adsorb hydrogen via Kubas interaction as well as the well-known van der Waals interaction. Finally, we construct a simple Monte-Catlo algorithm for Hydrogen uptake and show that Titanium metalloboranes in a MOF5 structure can absorb up to 11.5% hydrogen per weight at 100 bar of pressure.

## I. Introduction

Due to the geographical distribution and supply limitation of fossil fuel resources and the increasing perceived negative impact of carbon dioxide emission in the environment, it is advantageous for the societies to take action in the search and implementation of alternative energy systems. Hydrogen is a clean energy carrier and a potential substitute in this regard. However, to use it at the fullest capacity one needs to find an optimal way of its storage [1]. Among different ways of storing hydrogen, molecular sdsorption and solid-state storage are considered more attractive for the purpose of on-board applications. Despite an extensive search of an optimal hydrogen storage system, efficient hydrogen storage still remains illusive. This is because that a good storage system must meet three criteria: high gravimetric density; good thermodynamics, i.e., hydrogen bonding to material should not be neither too strong nor too weak; and good kinetics for reactions upon uptake as well as release of hydrogen [2].

Yildrim et al. proposed that Kubas interactions [3], i.e. interaction between the d-orbital of transition metals (e.g. Ti, Sc or Y) and the sigma bonding of the hydrogen molecule, could effectively mediate the hydrogen uptake. This interaction is specifically stronger when carbon nanotube is decorated with Ti atoms [4-6]. However, the drawback with this method is that transition metal (TM) on nanostructures tends to coalesce thus destroying the storage capacity. Using a first principle study, Singh et al. [7] showed that substitution of organic chain $C_6H_6$ in metal organic framework (MOF 5) with closo-$B_6H_6$ leads to a promising storage. In this study the authors decorated the metallocarborane with Ti and Sc. Their calculations yield a hydrogen binding energy to mettallocarborane of 30 kJ/mol as the result of Kubas interaction [7]. This energy is in the ballpark of DOE target. This result is quite attractive because MOF5 absorbs hydrogen at low temperature and thus its binding energy to hydrogen drops to very low values at room temperatures [8-10]. Another finding of reference 6 is that TM metals do not aggregate on the metallocarborane which is the advantage over carbon nanostructures [7].

This brings up possibilities of using planar form of $B_6H_6$ in MOF5 structure. Gosh et al. in 2001 synthesized planar $B_6H_6$ sandwiched by complex Rhenium [11]. Also Alexandrova et al. by using quantum chemistry simulations showed formation of planar aromatic $Li_6B_6H_6$ molecule [12,13]. Their study shows that planar $Li_6B_6H_6$ case is more stable than the competing closo-$Li_6B_6H_6$ by 31.3 kcal/mol [12]. In that study, the formation signature of π orbital's, σ bonding orbital's and the aromaticity of this inorganic benzene ring were identified [12]. This means to stabilize planar $[B_6H_6]$, six electrons must be transferred into the ring

In this article we present a planar-$B_6H_6$ molecule that is sandwiched with two metals ($M_2B_6H_6$ with M = Ti, Sc, Al). We extensively studied hydrogen storage capacities of $M_2B_6H_6$. Our results indicate potential application of this molecule for hydrogen storage with gravimetric density of up to 13 wt% $H_2$. We also replaced benzene ring in MOF5 with $M_2B_6H_6$ and formed another metalloborane. Our first principles show a stable structure. Finally, we performed Monte Carlo simulations on the hydrogen uptake of this metalloborane that shows hydrogen storage capacity of approximately 12 wt.% $H_2$ at room temperature.

## II. Metalloboranes: iso-electronic benzene

Boron is unique in its bonding characteristics [14-17] and is one of the few compounds able to establish a three-center bonding scheme [18,19] as well as being able to mimic the bonding structure of carbon. This produces a very rich bonding behavior that has only recently been appreciated [19-25]. Boron in a six-member benzene ring configuration is considered unstable. because Boron in this configuration is very electron deficient. This means that $B_6H_6$ electronic structure can be stabilized via two mechanisms; (i) the molecule can reconfigure its structure, making use of three-center bonding; or (ii) it can acquire electrons from its environment. Which mechanism is favored depends on the specifics of the environment such as the electron affinities of the surrounding atoms. For example, in Figure 1 we show the two different structures of $B_6H_6$, the closo- and the planer. Without the presence of the cations, the closo- structure is the lowest in energy. However, in the presence of cations the planer structures can be the lowest in energy. This is one of the purposes of this study, the investigation of the effects of charge transfer on the stability of this molecules, i.e. $B_6H_6$, Below we show that $B_6H_6$ is iso-electronic with benzene.

## III. Methodology

We confirmed the stability of $B_6H_6$ molecules decorated with transition metal using the FreeON Gaussian based code suite for structural studies [26] and the Becke-Lee-Yang-Parr Hybrid Functional [27,28]. We further used first-principles density-functional theory (DFT) calculations with the generalized gradient approximation (GGA) and Vanderbilt ultrasoft pseudopotentials [29] with Perdew Wang (PW91) exchange-correlation functional [30], as implemented in the Quantum Espresso ab initio simulation package [31]. For DFT calculations a plane wave cut off of 50 Ry was used to obtain convergence of the ground state total energy. In all energy calculations the molecule was placed in a large box with a side length of 20 Å to avoid interactions by its images. Structural relaxation of atomic positions was carried out until the residual forces were less than 0.05 eV/Å. Gamma point total energy calculations were carried out throughout the study.

## IV. Results and Discussions

**Metalloboranes:** Figure 1. (a-b) show the relaxed geometry for $Sc_2B_6H_6$ from two different views, and Figure 1 (c) shows the competing structure closo-$Sc_2B_6H_6$. Our simulations indicates that planar-$Sc_2B_6H_6$ ($Ti_2B_6H_6$) is lower in energy than the closo counterpart by 1.3 eV (2.4 eV). However, we found that closo-$Al_2B_6H_6$ is lower in energy than the planar-$Al_2B_6H_6$. In Table I and II we show the bond lengths, binding energies and partial charges for the planer and closo $M_2B_6H_6$ molecular species for M=Al, Sc and Ti As is shown, $Ti_2B_6H_6$ is the most stable. This can be partially explained based on ionization energies (IE) of Al, Sc, and Ti and the degree of charge transfer of the metal ions. As we have indicated, the relative stability of the metalized boron hydrides depends on the relative energies of charge transfer. Table III shows ionization energies for metal atoms Al, Sc, and Ti [32]. It is known that Al has larger $2^{nd}$ and $3^{rd}$ ionization energies than Sc and Ti, therefore it is more difficult for Al to yield its electron(s) to the ring. To further explore this, we calculated the relative charges on each atom in the $M_2B_6H_6$ as well as amount of charge transferred by the metal atom in both the planar and closo-$M_2B_6H_6$ configurations, respectively (Table II). As shown Al (Ti) has the lowest (largest) charge transfer in both planar and closo-structures, as well as having the longest (shortest) boron-boron bond length. Interestingly and noteworthy, Alexandrova et al., using the hybrid method at the MP2 level of theory also showed that planar $[B_6H_6]^{6-}$ is more stable than the highly aromatic closo-$[B_6H_6]^{2-}$ [12]. In Figure 2 we show the electron localization function (ELF) at an isosurface level of 0.63 for $Sc_2B_6H_6$. It is evident that the boron-boron bonds are due to the π-orbital aromatic bonds, as has been previously explored (Note $Al_2B_6H_6$ and $Ti_2B_6H_6$ show similar ELF and are not shown here). To summarize, we have shown that $Ti_2B_6H_6$ is the lowest energy structure and is stable due to the large charge transfer of the Titanium cation to the boron ring.

**Hydrogen Absorption:** Tables IV and V shows calculated binding energies for hydrogen absorbed onto $Sc_2B_6H_6$ and $Ti_2H_6B_6$ metalloboranes. We define the binding energy as the energy needed to remove an excess hydrogen molecule, where we also calculate and correct for the zero point energy of the hydrogen,

$$E_{BE} = \begin{cases} E(n,n) - E(n,n-1) + \frac{3}{2}\hbar\omega_0 \\ E(n,n+1) - E(n,n) + \frac{3}{2}\hbar\omega_0 \end{cases} \quad (1)$$

All of the binding energies are within 5 kJ/mole of 25 kJ/mole (including zero point energy, ZPE) that is the ideal thermodynamic range for efficient absorption and release of hydrogen for use in hydrogen storage. In Figure 3 shows the differential isosurface plot of the bonding charge density in $Sc_2B_6H_6$-(6,2)$H_2$ configuration. This configuration refers to the hydrated $Sc_2B_6H_6$ with 8 $H_2$ molecules attached to Sc on both sides of $Sc_2B_6H_6$. Charge accumulations (red) and depletion (blue) are shown between Sc and $H_2$ molecules. The H-H bond length is elongated from 0.74 Angstroms to 0.80 Angstroms due to the Kubas interaction. As is apparent, the boding is d-shell in character that is in accord to the Kubas mechanism. We also find that approximately 0.1 electrons are involved in the Kubas bond, which is in accord with the magnitude of the Kubas binding energy. To further illustrate the Kubas mechanism in Figure 4 we show the change in the density of state (DOS) of $Sc_2B_6H_6$ as the molecule absorbs hydrogen. As can be seen in the figure panels, the d-orbitals, which are above the Fermi energy (zero) are pushed down below the Fermi level as $H_2$ is absorbed to the molecule. It should be noted that these compounds do not suffer from the Titanium aggregation or the hydrogen dissociation problem that titanium decorated graphite materials suffer from [4,33]. This is because of the large charge transfer on the Titanium cation inhibits both aggregation and depopulation of the p-orbitals, which inhibits hydrogen dissociation.

**Molecular Organic Framework:** In Figure 5 shows the crystal structure of a Molecular Organic Framework (MOF). Specifically, we used MOF5 as prototype where we replaced the Terephthalic acid linker benzene ring with the $Ti_2B_6H_6$ molecule. We have calculated that this crystal structure has a lattice constant of 25.7 Angstroms. We also used this to calculate the excluded volume due to the molecular structure, which we have found to be approximately 20% of the total volume. We use this structure as a model for the chemical absorption for the calculation of the hydrogen uptake of this proposed system [10,34,35].

**Hydrogen Uptake in Molecular Organic Frameworks:** Using the results for hydrogen binding to Titanium Metalloboranes, and a modified Metropolis Monte-Carlo algorithm we have calculated total hydrogen uptake for Titanium Metelloborane-MOF5 system [36]. Specifically, we constructed a two-system model, Hydrogen in the gas phase and Hydrogen absorb to the

Titanium Metalloborane. Monte-Carlo algorithms is used to sample thermodynamically the systems states. For the hydrogen gas phase we use the equation of state developed by Lemmon et. al. [37]. This allows us to directly access the gas pressure of the system as the simulation progresses. We represent the state of the hydrogen gas by first defining the compressibility factor,

$$Z(n,P,T) = \frac{PV^*}{nkT} = 1 + \sum_{i=1}^{9} a_i \left(\frac{100K}{T}\right)^{b_i} \left(\frac{P}{1000\ bars}\right)^{c_i} \quad (2)$$

Where the constants are defined in Table I of reference [37], and $V^*$ is the total volume minus the excluded volume due to the MOF5 crystal. We calculate $V^*$ for Titanium metalloborane-MOF5 structure to be approximately 81% of the total volume. To proceed, we define a unit cell with the Ti2B6H6-MOF5 structure. For the unit cell depicted in Figure 5, there are 48 titanium atoms that have 240 binding sites. We calculate the binding energy with ZPE per titanium atom, which we show in Table IV. To start the calculation we first specify the number of Hydrogen molecules in the simulation and place them randomly at the 240 binding sites. We then choose a hydrogen molecule at random. If the Hydrogen molecule is in the gas phase, we randomly chose an open binding site and calculate the probability of the transition. If the atom is in the binding phase, we calculate the probability of it transitioning to the gas phase. The probabilities are given by,

$$P[i \to f] = \max\left[1,\ \text{Exp}[-\Delta E/kT]\right] \quad (3)$$

Where

$$E_{abs}(n,T) = \tfrac{3}{2}nkT + \sum_{i=1}^{n} E_{BE}^{i}$$
$$E_{gas}(n,P,T) = \tfrac{5}{2}nkT\, Z(n,P,T) \quad (4)$$

and

$$\Delta E = \begin{cases} \left(E_{abs}(n-1,T) + E_{gas}(n+1,P^*,T)\right) - \left(E_{abs}(n,T) + E_{gas}(n,P,T)\right) & \text{absorbate} \rightarrow \text{gas} \\ \left(E_{abs}(n+1,T) + E_{gas}(n-1,P^*,T)\right) - \left(E_{abs}(n,T) + E_{gas}(n,P,T)\right) & \text{gas} \rightarrow \text{absorbate} \end{cases}$$

From this simulation we obtain the pressure from the gas Equation and average it over ten million instances. Figure 6 shows our results for the density pressure curve of the simulated system for the temperatures 300 Kelvin and 500 Kelvin, as well as experimental data for comparison from reference [38]. These results show that for the aforementioned system, hydrogen storage up to 11.5% per weight at 100 bars and 300K is possible.

## V. Conclusions

Using first principles calculations, we have illuminated a new class of compounds, metalloboranes, which can be used for high capacity hydrogen storage. Metalloboranes are transition metal (TM) and borane compounds that obey a novel-bonding scheme. We have found that the transition metal atoms can bind up to 10 $H_2$-molecules with an average binding energy of 30 kJ/mole of $H_2$, which lies favorably within the reversible adsorption range. Among the first row Transition Metal (TM) atoms, Sc and Ti are found to be the optimum in maximizing the $H_2$ storage on the metalloborane cluster. Additionally, being ionically bonded to the borane molecule, the TMs do not suffer from the aggregation problem, which has been the biggest hurdle for the success of TM-decorated graphitic materials for hydrogen storage. Furthermore, we showed that since the borane 6-atom ring has identical bonding properties as carbon rings it is possible link the metalloboranes into metal organic frameworks, which are thus able to adsorb hydrogen via Kubas interaction as well as the well-known van der Waals interaction. Finally, we constructed a simple Monte-Catlo algorithm for Hydrogen uptake and showed that Titanium metalloboranes in a MOF5 structure can absorb up to 11.5% hydrogen per weight at 100 bar of pressure, which if realized industrially, would be a significant advance in hydrogen storage.

## VI. Acknowledgments

The authors would like to acknowledge the support given by Welch Foundation (Grant J-1675), the ARO (Grant W911Nf-13-1-0162), the Texas Southern University High Performance Computing Center (http:/hpcc.tsu.edu/; Grant PHY-1126251) and NSF-CREST CRCN project (Grant HRD-1137732). Patent: "Metalloboranes for High Density Hydrogen Storage"; D7291; U.S. Serial No.: 62/031, 993

## VII. List of Tables

**Table I:** Bond lengths (Angstroms) and formation energies of different species of $M_2B_6H_6$ in the planar and closo configurations (M=Al, Sc, Ti) as calculated using the Becke-Lee-Yang-Parr Hybrid Functional and the 6-31G** basis set in the FreeON code suite.

| Species | Formation Energy (eV) | Gap (eV) | Bond Lengths | | |
|---|---|---|---|---|---|
| | | | B-H (A) | B-M (A) | B-B (A) |
| closo-$Al_2B_6H_6$ | -53.07 | 4.109 | 1.191 | 2.411 | 1.733 |
| planer-$Al_2B_6H_6$ | -49.73 | 1.423 | 1.177 | 2.155 | 1.766 |
| closo-$Sc_2B_6H_6$ | -51.40 | 1.524 | 1.194 | 2.496 | 1.740 |
| planer-$Sc_2B_6H_6$ | -52.16 | 2.476 | 1.191 | 2.378 | 1.711 |
| closo-$Ti_2B_6H_6$ | -52.89 | 1.094 | 1.201 | 2.284 | 1.733 |
| planer-$Ti_2B_6H_6$ | -55.66 | 1.967 | 1.189 | 2.249 | 1.695 |

**Table II:** Partial charges on the different species of $M_2B_6H_6$ in the planar and closo configurations (M=Al, Sc, Ti).

| Species | Charge M | Charge B | Charge H |
|---|---|---|---|
| closo-$Al_2B_6H_6$ | 1.075 | -0.325 | -0.032 |
| planer-$Al_2B_6H_6$ | 2.387 | -0.787 | -0.008 |
| closo-$Sc_2B_6H_6$ | 1.138 | -0.345 | -0.033 |
| planer-$Sc_2B_6H_6$ | 2.499 | -0.820 | -0.012 |
| closo-$Ti_2B_6H_6$ | 1.789 | -0.559 | -0.037 |
| planer-$Ti_2B_6H_6$ | 2.663 | -0.855 | -0.032 |

**Table III:** Ionization Energies from Reference [32]

| Species | $1^{st}$ Ionization (eV) | $2^{nd}$ Ionization (eV) | $3^{rd}$ Ionization (eV) |
|---|---|---|---|
| Aluminum | 0.598 | 1.882 | 2.844 |
| Scandium | 0.656 | 1.279 | 2.475 |
| Titanium | 0.683 | 1.357 | 2.748 |

**Table IV:** Binding Energies for $Sc_2B_6H_6+(n,m)H_2$

| Number of Hydrogen | Hydrogen Configuration | Binding Energy (kJ/mole) |
|---|---|---|
| 1 | (0,0)→(0,1) | 30.02 |
| 2 | (0,1)→(1,1) | 35.06 |
| 3 | (1,1)→(1,2) | 21.72 |
| 4 | (1,2)→(2,2) | 34.07 |
| 5 | (2,2)→(2,3) | 30.19 |
| 6 | (2,3)→(3,3) | 37.85 |
| 7 | (3,3)→(3,4) | 13.63 |
| 8 | (3,4)→(4,4) | 14.34 |

**Table V:** Binding Energies for $Ti_2B_6H_6+(n,m)H_2$

| Number of Hydrogen | Hydrogen Configuration | Zero Point Energy of $H_2$ (kJ/mole) | Binding Energy (kJ/mole) |
|---|---|---|---|
| 1 | (0,0)→(0,1) | 4.22 | 25.79 |
| 2 | (0,1)→(1,1) | 4.24 | 23.45 |
| 3 | (1,1)→(1,2) | 4.25 | 26.12 |
| 4 | (1,2)→(2,2) | 4.27 | 28.80 |
| 5 | (2,2)→(2,3) | 4.28 | 23.19 |
| 6 | (2,3)→(3,3) | 4.28 | 22.73 |
| 7 | (3,3)→(3,4) | 4.30 | 20.13 |
| 8 | (3,4)→(4,4) | 4.31 | 22.01 |
| 9 | (4,4)→(4,5) | 4.40 | 17.89 |
| 10 | (4,5)→(5,5) | 4.50 | 17.75 |

## VIII. List of Figures

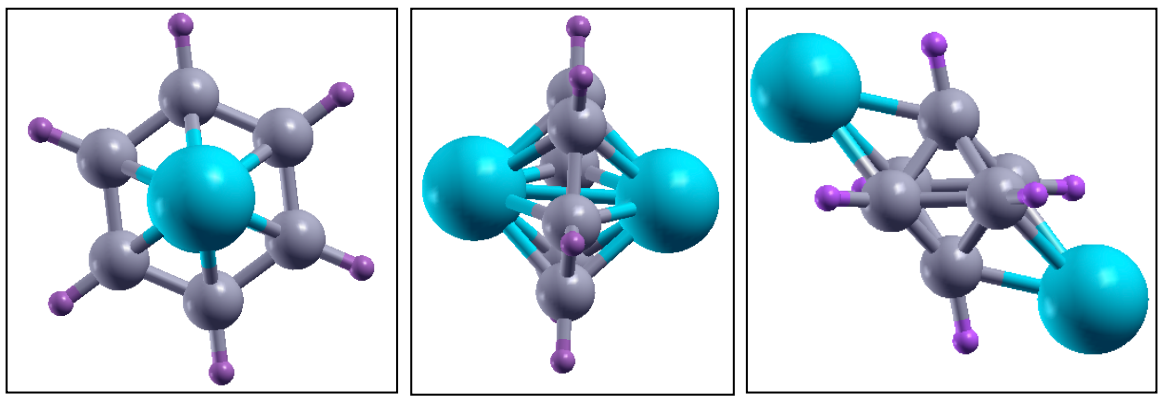

**Figure 1:** Planar M₂B₆H₆ (M=Al, Sc, Ti) (a) top view planar, (b) side views planar, and (c) closo-, respectively. Symbols represent B (gray), Sc (large blue), and H (purple), respectively.

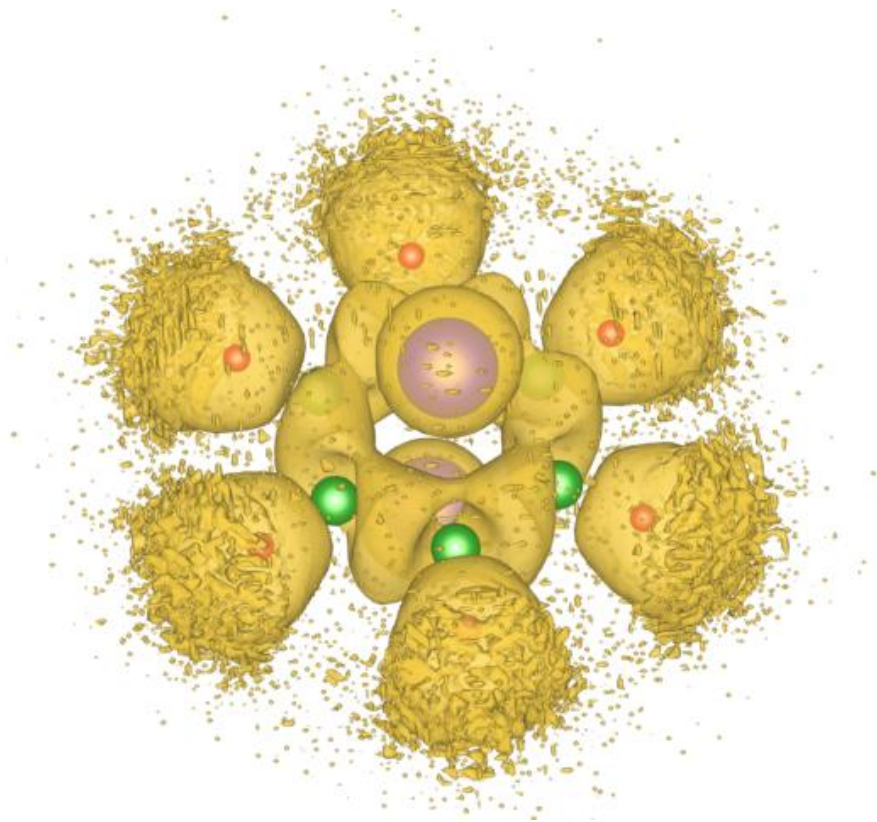

**Figure 2:** The electron localization function (ELF) at isosurface level 0.63 for $Sc_2B_6H_6$ computed in Quantum Espresso. As can be seen the six-boron ring has a nicely formed Pi-orbital above and below the boron ring. Figure shows H atoms in red, B atoms in green and Sc atoms in purple [39].

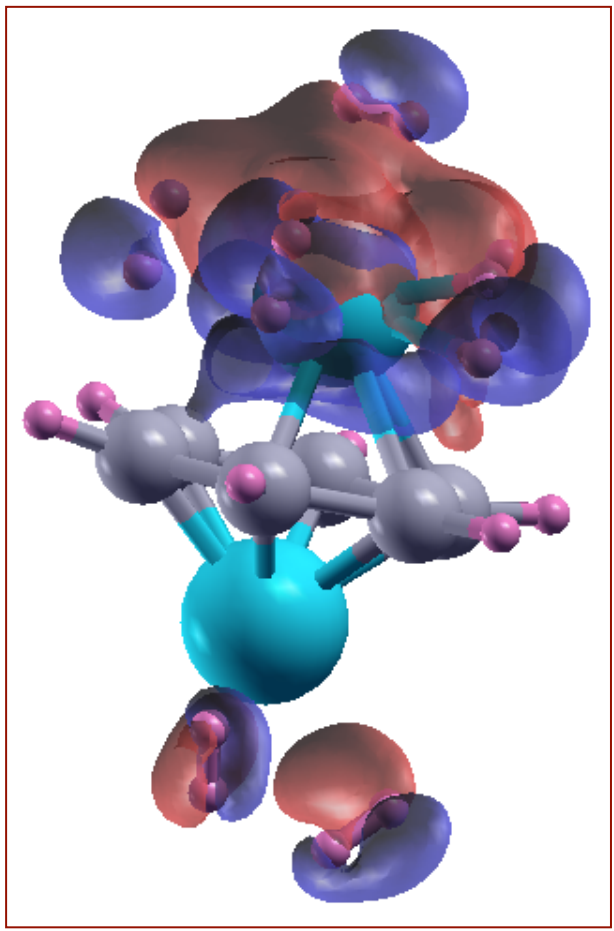

**Figure 3**: Isosurface bonding charge density in $Sc_2B_6H_6$-(6,2)$H_2$ configuration. This configuration refers to hydrated $Sc_2B_6H_6$ with 8 $H_2$ molecules attached to Sc on both sides of $Sc_2B_6H_6$. Charge accumulations (red) and depletion (blue) between Sc and $H_2$ molecules. The H-H bond length is enlogated from 0.74 A to 0.80 A due to the Kubas interaction. Generated using XCrystalGen [40].

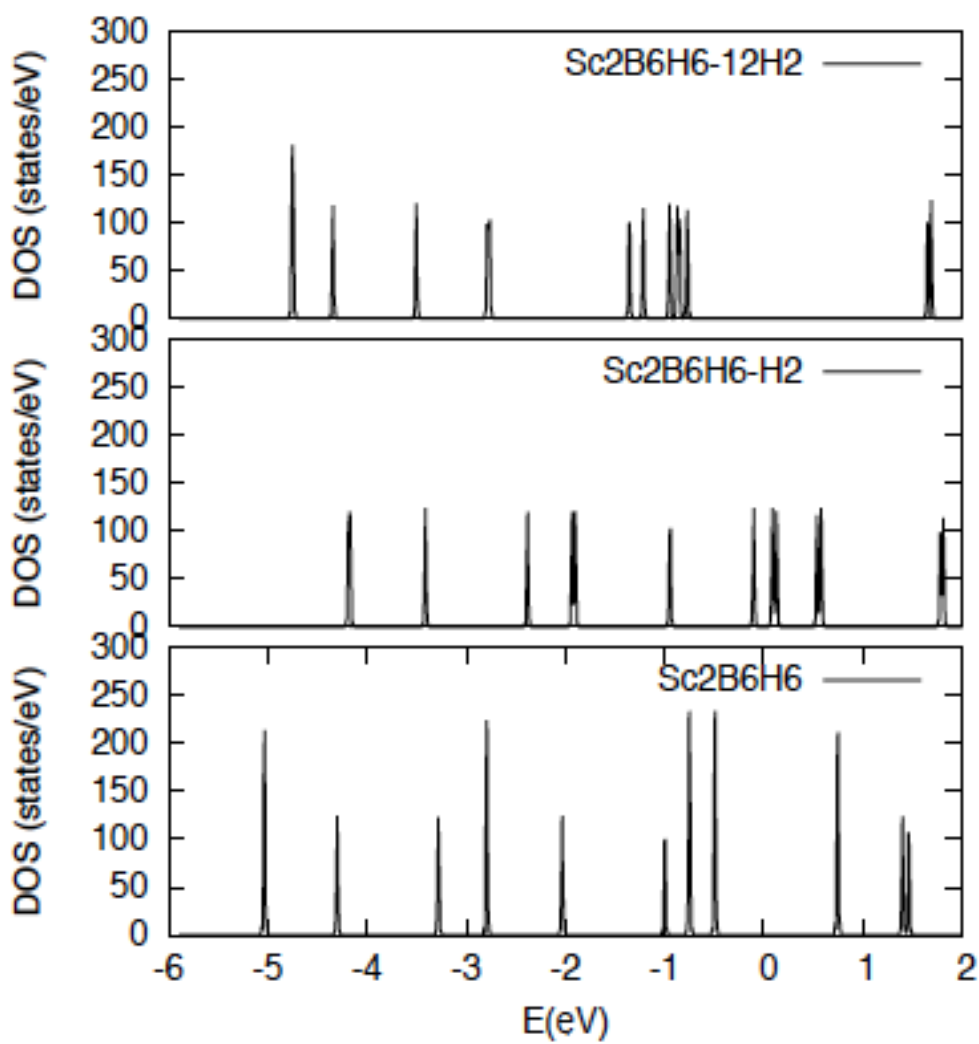

**Figure 4:** Density of states (DOS) for: (bottom panel) $Sc_2B_6H_6$; (middle panel) $Sc_2B_6H_6$-$(1,0)H_2$; and (top panel) $Sc_2B_6H_6$-$(6,6)H_2$ configurations, respectively. Note that due to Kubas interaction between $H_2$ s orbitals and the d-oribtals of Sc. It is shown that d orbitals, which are above the Fermi energy (zero) are pushed down below the Fermi level as $H_2$ is absorbed to the molecule.

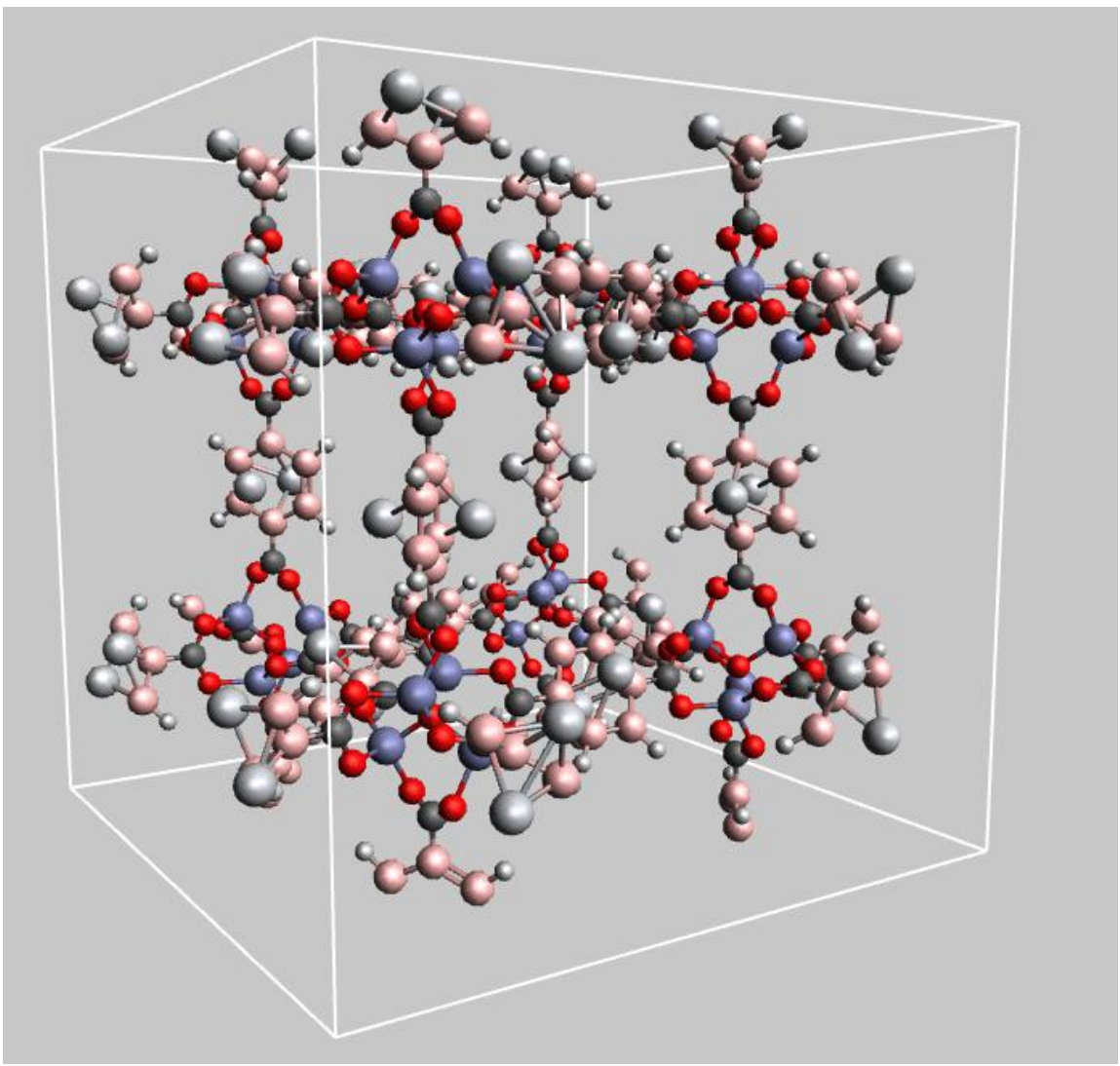

**Figure 5:** Ti$_2$B$_6$H$_4$(COOH)$_2$ in a metal organic framework (MOF5). The addition of the titanium to the MOF5 system significantly increase hydrogen uptake at 300K

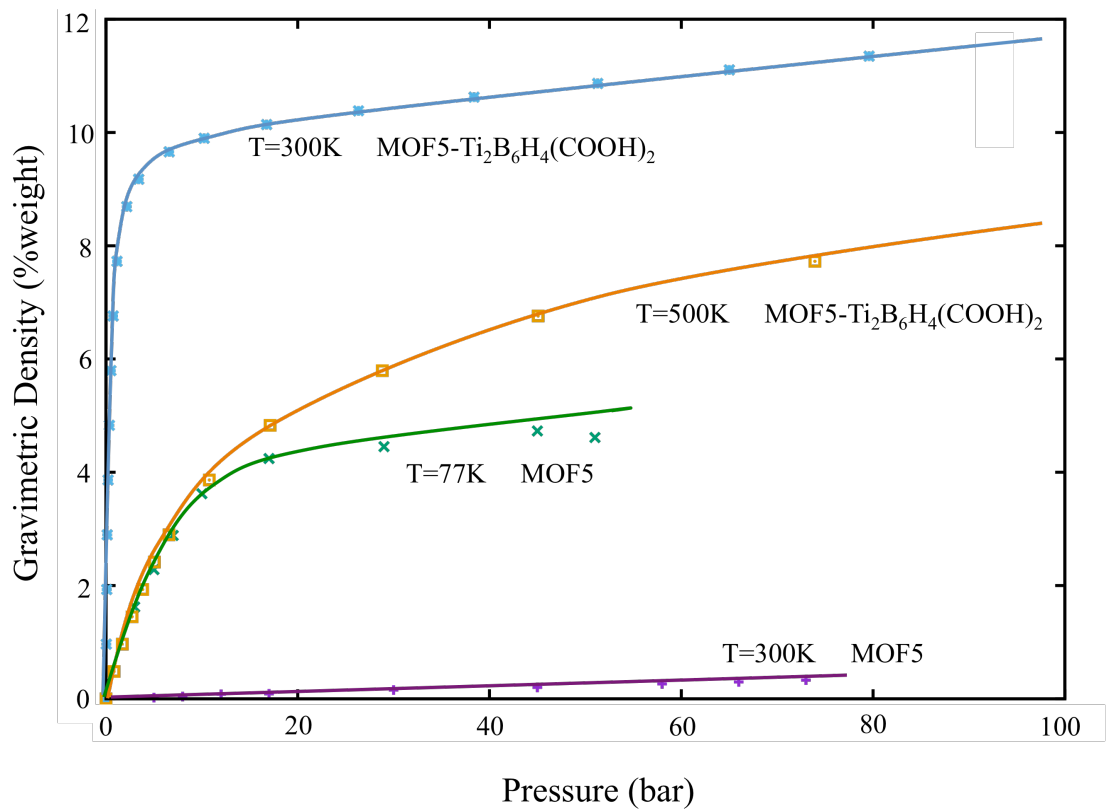

**Figure 6:** Absorption vs. pressure curve for the MOF5-Ti$_2$B$_6$H$_4$(COOH)$_2$ system computed using the Hydrogen Binding energies from Table V and a Monte Carlo algorithm. The blue curve is computed at 300K (room temperature) and the orange curve at 500K (auto exhaust temperature). For comparison we show experimental data at 77K (green curve) and 300K (purple curve) from reference [38]